# Performance Evaluation of Wi-Fi comparison with WiMAX Networks

[1]M.Sreerama Murty, [2] D.Veeraiah, [3]A.Srinivas Rao

[1]Department of Computer Science and Engineering
Sai Spurthi Institute of Technology,Khammam,Andhra Pradesh,India
`sreerammaturi@yahoo.com`
[2]Department of Computer Science and Engineering
Sai Spurthi Institute of Technology,Khamamm,Andhra Pradesh,India
`veeraiahdvc@gmail.com`
[3]Department of Computer Science and Engineering
Sai Spurthi Institute of Technology,Khamamm,Andhra Pradesh,India
`srinivas.ada@gmail.com`

*Abstract*

*Wireless networking has become an important area of research in academic and industry. The main objectives of this paper is to gain in-depth knowledge about the Wi-Fi- WiMAX technology and how it works and understand the problems about the WiFi- WiMAX technology in maintaining and deployment. The challenges in wireless networks include issues like security, seamless handover, location and emergency services, cooperation, and QoS.The performance of the WiMAX is better than the Wi-Fi and also it provide the good response in the access. It's evaluated the Quality of Service (Qos) in Wi-Fi compare with WiMAX and provides the various kinds of security Mechanisms. Authentication to verify the identity of the authorized communicating client stations. Confidentiality (Privacy) to secure that the wirelessly conveyed information will remain private and protected. Take necessary actions and configurations that are needed in order to deploy Wi-Fi -WiMAX with increased levels of security and privacy*

## 1. Introduction

Recently wireless networking has become an important area of research in academia and industry. This is due to the huge diversity of wireless network types, which range from Wireless Fidelity network (Wi-Fi) covering smallest area to Worldwide Interoperability for Microwave Access networks (Wimax) covering up to several miles. All these typesof networks have been developed separately with different usage and applications scenarios, which make networking between them a challenging task.

### 1.1 Wi-Fi Network

Wi-Fi technology builds on IEEE 802.11 standards and Wi-Fi technology is still using local area network (LAN) for the predictable future. Wi-Fi can be used as various handheld devices. The handheld devices are connected to internet by using the connection of Wi-Fi. The access of the Wi-Fi network is limited to a specific area and should not expend the network. This network is only for within the specified area only. And its established limited in some restricted place.

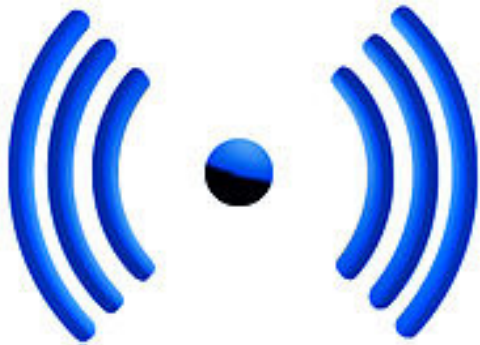

**Figure 1.1 Wi-Fi Signal Logo**

Wi-Fi is not a technical term. However,. Routers that incorporate a digital subscriber line modem or a cable modem and a Wi-Fi access point, often set up in homes and other premises, can provide Internet access and internetworking to all devices connected (wirelessly or by cable) to them. The connection was established from one system to another directly without any intermediate node. This mode of connection is known as ad-hoc network. The connection establishment of the Wi-Fi is using some consumer electronic devices.

The Wi-Fi technology used different ways as follows

- **City-wide Wi-Fi**

This type of network implementation only for the city-wide network connection. It's performance is not gives the better response to the established network, so this type of network application has been canceled.

- **Campus-wide Wi-Fi**

Many traditional college campuses provide at least partial wireless Wi-Fi Internet coverage and also major university to offer completely wireless Internet access across the entire campus.

- **Internet Access**

The access of an internet using the Wi-Fi network using the handheld devices. The connection was indicated as access points, by using this access point internet is work. Routers that incorporate a digital subscriber line modem or a cable modem and a Wi-Fi access point, often set up in homes and other premises, can provide Internet access and internetworking to all devices connected (wirelessly or by cable) to them.

## 1.2  WiMax Network

WiMAX (Worldwide Interoperability for Microwave Access) is a telecommunications protocol that provides fixed and mobile Internet access. The WiMAX produce up to 40 Mbit/s using the IEEE 802.16m and also release the maximum speed is up to 1 Gbit/s . The name "WiMAX" was created by the WiMAX Forum. The forum describes WiMAX as "a standards-based technology enabling the delivery of last mile wireless broadband access as an alternative to cable and DSL".

Clarifications of the formal names are as follow:

- 802.16d is refer to as Fixed Wimax,it's not support the mobility of a network

- 802.16e referred as the Mobile Wimax, it was establishes as the wireless network, and also called Mobile Network

The bandwidth and range of WiMAX make it suitable for the following potential applications:

i. Providing a wireless alternative to cable and DSL for "last mile" broadband access.

ii. Providing a source of Internet connectivity as part of a business continuity plan.

The following Objectives are used for WiMAX Networks.

- Broadband
- Backhaul
- Triple-play
- Rapid deployment

## 2. Literature Survey

The paper title "**Technology-integration frame work for fast and low cost handover, case study: WIFI-WIMAX Network**" .The growth of wireless communication has been, in a few years, important thanks to the advantages they offer such as deployment flexibility and user mobility during communications. Several wireless technologies have emerged.

The paper title "*WiFi and WiMAX secure deployments*" the security intrusion that may occur during handover is discussed. The demand for higher data rates, different modulations and frequency transmissions, improved Quality of Service (QoS), enhanced security and authentication mechanisms. When the technology was brought to the market, there were concerns if products from different vendors could meet interoperability.

The IEEE 802.16 currently employs the most sophisticated technology solutions in the wireless world, and correspondingly it guarantees performance in terms of covered area, bit-rate, and QoS.

WiMAX implements stronger security mechanisms and succeeds to block most of the threats in a wireless network. Nevertheless some weaknesses still exist in WiMAX as well; in the following, we will try to identify the recommendations for WiFi and WiMAX, on how specific mechanisms should be used, how specific security options shall be set and if new security mechanisms, additional to the ones available with Wi-Fi and WiMAX, are needed in order for the network will operate more securely and robustly.

In the paper entitled "**A multi-standardfrequency offset synchronization scheme for 802.11n, 802.16d,LTE and DVB-T/H systems**" by J. González-Bay ´on et al.carrier frequency offset in OFDM systems is discussed wherecommon synchronization structure for all these systems is proposed

In the samearea P. T. T. Pham and T. Wada's paper "*Effective schemeof channel tracking and estimation for mobile WiMAX DLPUSCSystem*" discussed the packet error rate (PER) and user throughput in various channels.

K.-P. Lin and H.-Y. Wei discussed a new random walk mobility model in "*Paging and location management in IEEE802.16j multi hop relay network*". The proposed model is suitable for multi hop relay network, where the handover process is frequently performed Finally, "*Multi mode flex-inter leaver core for baseband processor platform*" by R. Asghar and D Liu

323

introduces a new flexible interleave architecture supporting many standards like WLAN, WiMAX, HSPA+, LTE, and DVB at the system level. Both maximum flexibility and fast switch ability were examined during run time.

## 3. Analysis of Wi-Fi and WiMAX

### 3.1 WiMAX technology

It's standards for worldwide interoperability for Microwave Access and also known as 802.16.it was designed for the longer range of wireless network connections such as to provide internet access to a particular geographic area. It can be established the range from 39 miles to 6 miles to 30miles.

WiMAX technology is a standard based wireless technology which is used to provide internet access and multimedia services at very high speed to the end user .The current WiMAX revision is based upon IEEE Std 802.16e-2005 IEEE 802.16e-2005 improves upon IEEE 802.16-2004 by:

- Advanced antenna diversity schemes, and hybrid automatic repeat-request (HARQ)
- Adaptive Antenna Systems (AAS) and MIMO technology
- Denser sub-channelization, thereby improving indoor penetration
- Introducing Turbo Coding and Low-Density Parity Check (LDPC)
- Introducing downlink sub-channelization, allowing administrators to trade coverage for capacity or vice versa

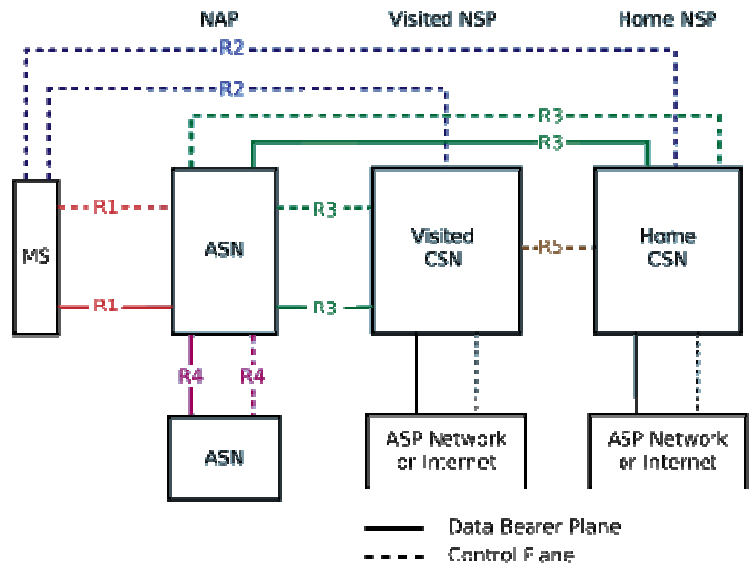

**Figure 3.1 WiMAX Architecture**

The WiMAX architecture was connected with an IP network, it was connected by the network service providers i.e. ISP (Internet Service Providers).it's designed seamless integration capability of other networks with an packets sending in ad-hoc mode.

The WiMAX forum proposal defines a number of components, plus some of the interconnections (or reference points) between these, labeled R1 to R5 and R8:

- SS/MS: the Subscriber Station/Mobile Station
- ASN: the Access Service Network[20]
- BS: Base station, part of the ASN
- ASN-GW: the ASN Gateway, part of the ASN
- CSN: the Connectivity Service Network
- HA: Home Agent, part of the CSN
- AAA: Authentication, Authorization and Accounting Server, part of the CSN
- NAP: a Network Access Provider
- NSP: a Network Service Provider

### 3.1.2 Mobility

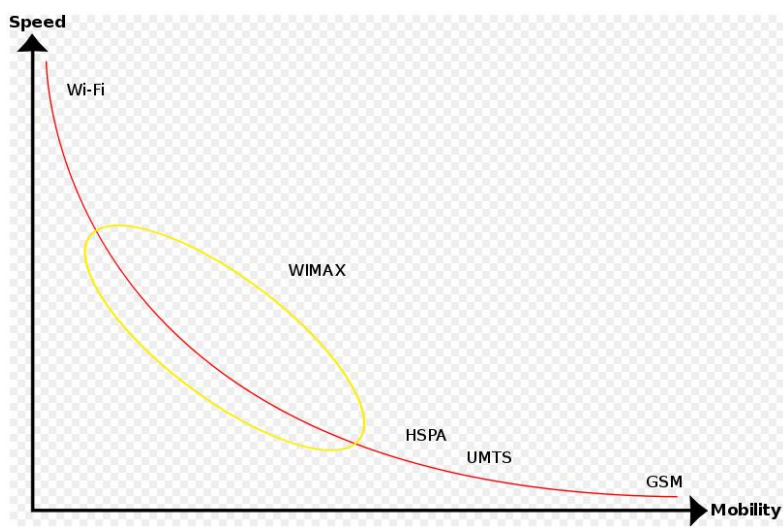

**Figure 3.2 Speed vs. Mobility of Wireless systems: Wimax, HSPA, UMTS, and GSM**

The comparison of speed and mobility of Wimax, HSPA, UMTS, and GSM drastically increase from Wi-Fi to GSM. The wireless systems speed is changed as per the no of access points in a network.

### 3.2 Wi-Fi technology

This technology is used in LAN; it allows the connection using wireless devices. The entire network was established with in the small amount of area like rooms. It could not run the outside environment. Because of the range of the bandwidth and frequency is limited to access the networks.

Wi-Fi can be used the handheld devices like laptops and pc's. The connection was established by the network adapters. Spaces where cables cannot be run, such as outdoor areas and historical buildings, can host wireless LANs. The price of chipsets for Wi-Fi continues to drop, making it an economical networking option included in even more devices. Wi-Fi has become widespread in corporate infrastructures. Products designated as "Wi-Fi Certified" by the Wi-Fi Alliance are backwards compatible. New protocols for quality-of-service (WMM) make Wi-Fi more suitable for latency-sensitive applications (such as voice and video); and power saving mechanisms (WMM Power Save) improve battery operation.

Wi-Fi is a family of networking, it sometimes called Ethernet. The Wi-Fi is used 802.11 protocols
Standards for wireless network. The speed and Wi-Fi network difference factors like freq, bandwidth. Generally Wi-Fi is designed for the medium range data transfers i.e. 100 to 300 feet in indoor.

The following versions are used as the Wi-Fi

| Version | Description |
|---|---|
| 802.11b | It support data transfer rate of 11 Mbps |
| 802.11a | Data Transfer rate is 54Mbps,freq:5GHZ |
| 802.11g | Similar data transfer rate of 802.11a but freq:2.4GHZ |
| 802.11n | It supports speeds up to 5 times better then the 802.11g.it can use the either 2.4GHZ or 5GHZ frequency. |

### 3.2.1 Mobility

The mobility of a network is based on networking points. It's worked as the short-range wireless networking, such as to network PCs within the building. The following graph represents comparison of the speed and mobility of the network.

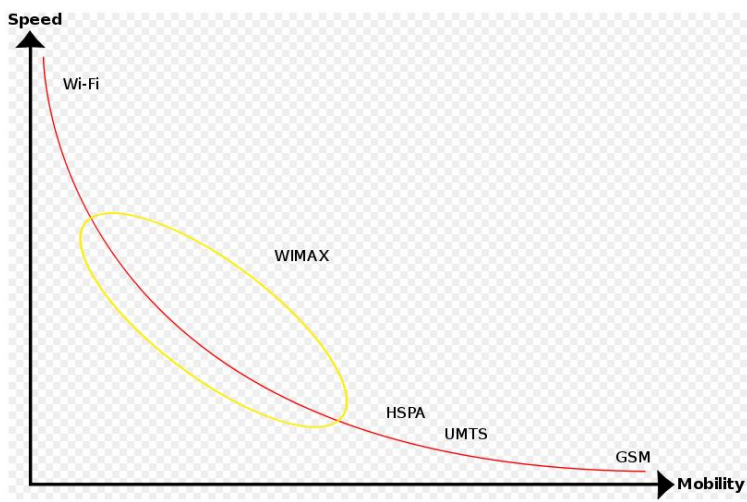

**Figure 3.2.1 Speed vs Mobility of Wireless systems:wifi,HSPA,UMTS,GSM**

## 4. Comparisions of WiMAX and WiFi

- The WiMax network is to establish by any network service providers. and also used in LAN

- WiMAX network execute a connection oriented MAC while Wi-Fi runs on the CSMA/CA protocol, which is wireless and strife based

- WiMAX is faster than the Wi-Fi, because is type of the connection in that area.

- The major difference of the WiMAX and Wi-Fi is speed and distance of a network

- The QoS of the both the networks are simple and reliable.

The following major comparisons are involved the Wimax and Wi-Fi:

| Wi-Fi | WiMAX |
| --- | --- |
| Connection Oriented | Connection Less |
| Limited area | Depends on the Networks establishments |
| Use the versions 802.11b,802.11a,802.11g,802.11n | Use the versions 802.16 |
| Less bandwidth | Medium Band width |
| Limited access points | No of access points |
| Connection must be reliable | Connect is Unreliable |

### 4.1 Technical Comparisons of Wi-Fi and WiMAX

The following data is compare the both Wi-Fi and WiMAX

| Standard | Familiy | Primary Use | Downlink (Mbit/s) | Uplink (Mbit/s) | Description |
| --- | --- | --- | --- | --- | --- |
| Wimax | 802.16 | Mobile Internet | 128 (in 20MHz bandwidth) | 56 (in 20MHz bandwidth) | WiMAX update IEEE 802.16mexpected to offer peak rates of at least 1 Gbit/s fixed speeds and 100Mbit/s to mobile users |
| Wi-Fi | 802.11 | Mobile Internet | 300 (using 4x4 configuration in 20MHz bandwidth) or 600 (using 4x4 configuration in 40MHz bandwidth | | Antenna, RF front end enhancements and minor protocol timer tweaks have helped deploy |

| | | | | | long range P2Pnetworks compromising on radial coverage, throughput and/or spectra efficiency (310km & 382km) |
|---|---|---|---|---|---|

Table 4.1 comparisons of Wi-Fi and Wimax

## 5. Conclusion

The performance of Wi-Fi compared with of WiMAX is good response of a wireless network. The problems in Wi-Fi network is overcome by the WiMAX network. Here the enter problem of the Wi-Fi network is restricted area. But the WiMAX has no restriction to work. Both of the networks are reliable networks. Compare with Wi-Fi network and WiMAX technology is more secure, reliable service.

**References:**

[1]. L.M.S.C. of the IEEE Computer Society, "Wireless LAN Medium Access Control (MAC) and Physical Layer (PHY) specifications: Higher-Speed Physical Layer Extension in the 2.4 GHz Band," ANSI/IEEE Standard 802.11-1999TM.

[2]. L.M.S.C. of the IEEE Computer Society, "Wireless LAN Medium Access Control (MAC) and Physical Layer (PHY) specifications," Amendment 6: Medium Access Control (MAC) Security Enhancements. IEEE Standard 802.11gTM- 2003.

[3]. WiMAX Forum, "Fixed, nomadic, portable and mobileapplications for 802.16-2004 and 802.16e WiMAX networks,"November 2005.

[4]. S. Fluhrer, I. Martin, and A. Shamir, "Weaknesses in the keyscheduling algorithm of RC4," in Proceedings of the 8th AnnualWorkshop on Selected Areas in Cryptography, Toronto, Canada,August 2001.

[5]. WiMAX Forum, "WiMAX Forum Web page," September 2008, http://www.wimaxforum.org/.

[6]. LAN/MAN Standards Committee, "IEEE 802.11i: Amendment6: Medium Access Control (MAC) Security Enhancements,"IEEE Computer Society, Standard, April 2004.

[7]. LAN/MAN Standards Committee, "IEEE 802.11e Amendment8: Medium Access Control (MAC) Quality of ServiceEnhancements," IEEE Computer Society, Standard, November2005.

[8]. M. Kassab and J.-M. Bonnin, "Optimized layer-2 handoverinWiFi-WiMAX networks," Research Report, TelecomBretagne, 2009.

[9]. M. Kassab, A. Belghith, J.-M. Bonnin, and S. Sassi, "Fastand secure hanfoffs for 802.11 infrastructures networks,"NetCon05 Lannion France, november 2005.

[10]. I. L. S. Committee, "Part 16: Air interface for fixed broadbandwireless access systems," IEEE Computer Society, Standard,June 2004.

## Acknowledgements

I would like to thanks the authors' who are help to this paper and their valuable suggestions to improve the paper.

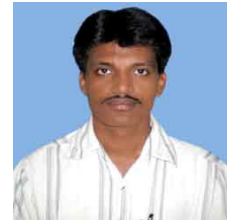

**M.SreeramaMurthy** ,M.Tech in CSE in  University College of Engineering ,JNTU,Kakinada.B.Tech in IT from JNTU,Hyderabad. And now presently working as Assistant Professor Sai Spurthi Institute of Technology,Khammam,Andhra Pradesh,India.His research interests includes Mobile Computing, Computer Networks ,Image Processing,DataMining and Embedded Systems.

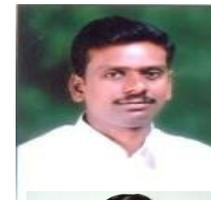

**D. Veeraiah** Received M.Tech in Computer Science and Engineering  from Anurag Engineering Coolge (JNTUH),B.Tech in Information Technology from Mother Teresa Institute of Science and Technology( JNTU,Hyderabad). And now presently working as Assoc. Professor Sai Spurthi Institute of Technology,Khammam.His research interests includes Image Processing,Computer Networks,Mobile Computing,Compiler Design.

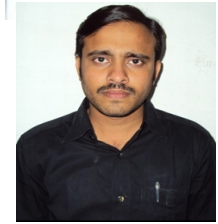

**A.Srinivas Rao** Received M.Tech in Computer Science and Engineering  from Anurag Engineering College  (JNTUH),B.Tech in Information Technology from Mother Teresa Institute of Science and Technology( JNTU,Hyderabad). And now presently working as Assoc. Professor Sai Spurthi Institute of Technology,Khammam.His research interests includes Mobile Computing,Image Processing,Computer Networks,Compiler Design